\documentclass[a4paper]{jpconf}
\usepackage{graphicx}
\usepackage{epstopdf}
\usepackage{amsmath}
\usepackage{hyperref}

\newcommand{\figref}[1]{Fig.~\ref{#1}}
\newcommand{\pypbec}{\texttt{PyPBEC}}
\newcommand{\python}{\texttt{Python}}

\newcommand{\Gup}{\ensuremath{\Gamma_\uparrow}}
\newcommand{\Gdown}{\ensuremath{\Gamma_\downarrow}}
\newcommand{\Am}{\ensuremath{A_m}}
\newcommand{\Em}{\ensuremath{E_m}}
\newcommand{\Lindblad}{\ensuremath{\mathcal{L}}}
\newcommand{\ddt}[1]{\ensuremath{\frac{{\rm d}{#1}}{{\rm d}t}}}

\begin{document}

\title{Phase transitions of light in a dye-filled microcavity: observations and simulations}

\author{R.A. Nyman, H.S. Dhar, J.D. Rodrigues and F. Mintert}

\address{Physics Department, Imperial College London, SW7 2AZ, UK}

\ead{r.nyman@imperial.ac.uk}

\begin{abstract}
Photon thermalisation and condensation in dye-filled microcavities is a growing area of scientific interest, at the intersection of photonics, quantum optics and statistical physics. We give here a short introduction to the topic, together with an explanation of some of our more important recent results. A key result across several projects is that we have a model based on a detailed physical description which has been used to accurately describe experimental observations. We present a new open-source package in \python{} called \pypbec\ which implements this model. The aim is to enable the reader to readily simulate and explore the physics of photon condensates themselves, so this article also includes a working example code which can be downloaded from the GitHub repository.
\end{abstract}

\section{Photon condensation in dye-filled microcavities}

Light in resonators which are filled with a fluorescent medium has been studied in many guises, with the most notable variant being the laser. More recently attention has turned to what happens when the medium also interacts strongly with the light, especially when the resonator supports multiple modes. 

When the coupling between light and matter is coherent, this is known as the strong-coupling regime of cavity quantum electrodynamics (CQED). The resulting hybrid light-matter particles are known as polaritons, and if pumped sufficiently strongly can cause lasing~\cite{Christopoulos07, KenaCohen10, Bajoni08}. The particles may thermalise among themselves to form polariton condensates, which has been shown in planar dielectric microcavities filled with inorganic semiconductors~\cite{Kasprzak06} or organic dyes~\cite{Plumhof14, Daskalakis14}, and also plasmon-polariton metamaterial systems~\cite{Ramezani17}.

There is a simpler, incoherent form of strong coupling between light and matter, which is reached when the absorption of light by the gain medium is at least as fast as loss from the resonator. This is the regime in which photon thermalisation to room temperature and Bose-Einstein condensation (BEC)~\cite{Klaers10} can occur, and this regime is the topic of this work. The basic principles have been described elsewhere, but we will start by re-stating them in a more general form. We then review some of our group's recent experimental and theoretical results, with the perspective that photon condensation is an example of a wider class of dynamical phase transitions, from which one can learn a lot about phase transitions in general. Finally, we present a new suite of open-source software that we have published to simulate photon condensation called \pypbec, including a result on the statistical properties of multi-mode microlasers derived using the software.

\subsection{How light reaches thermal equilibrium}

Optical resonators (and metamaterials) modify the density of states (DoS) available for light. We consider here a class of resonators which provide a large gap in the DoS, with several modes lying just at energies above the gap energy. Furthermore, the DoS increases with energy, such that there is exactly one lowest-energy state if the the system is finite in size. The simplest such resonators are open microcavities of length a few half-wavelengths of the light and radius of curvature at least a few tens of microns.

A fluorescent medium is introduced to interact with the light. The medium is chosen so that the absorption and emission spectra do not extend beyond the gap in the DoS for light. Thus the combination of resonator and medium provide a robust energy minimum for the light. They also overlap slightly, in that there is an energy range where both absorption and emission of light can occur rapidly. In that overlap region, we require that the ratio of absorption and emission spectra follows a Boltzmann factor: absorption dominates at higher energies, emission dominates at lower energies. This relation is valid for many fluorescent dyes (known as the Kennard-Stepanov relation), some doped laser glasses (known as the McCumber relation) and many semiconductors (a limit of the van Roosbroeck-Shockley relation).

If the resonator lifetime is longer than the absorption timescale, and most of the fluorescence light is emitted into the resonator, the average population of photons on the resonator modes depends on the ratio of absorption to emission. Since we chose the material for the Boltzmann-type relation between absorption and emission, the photon populations will follow a thermal factor. When the populations increase, stimulated emission means that the Bose-Einstein distribution is the valid distribution (i.e. with population controlled by the chemical potential). With sufficient population, the number in the lowest-energy state above the gap energy will diverge, while numbers of excited states saturate. Thus, Bose-Einstein condensation is observed.

\begin{figure}[h]
\centering
	\vspace{-2ex}
		\includegraphics[width=50mm]{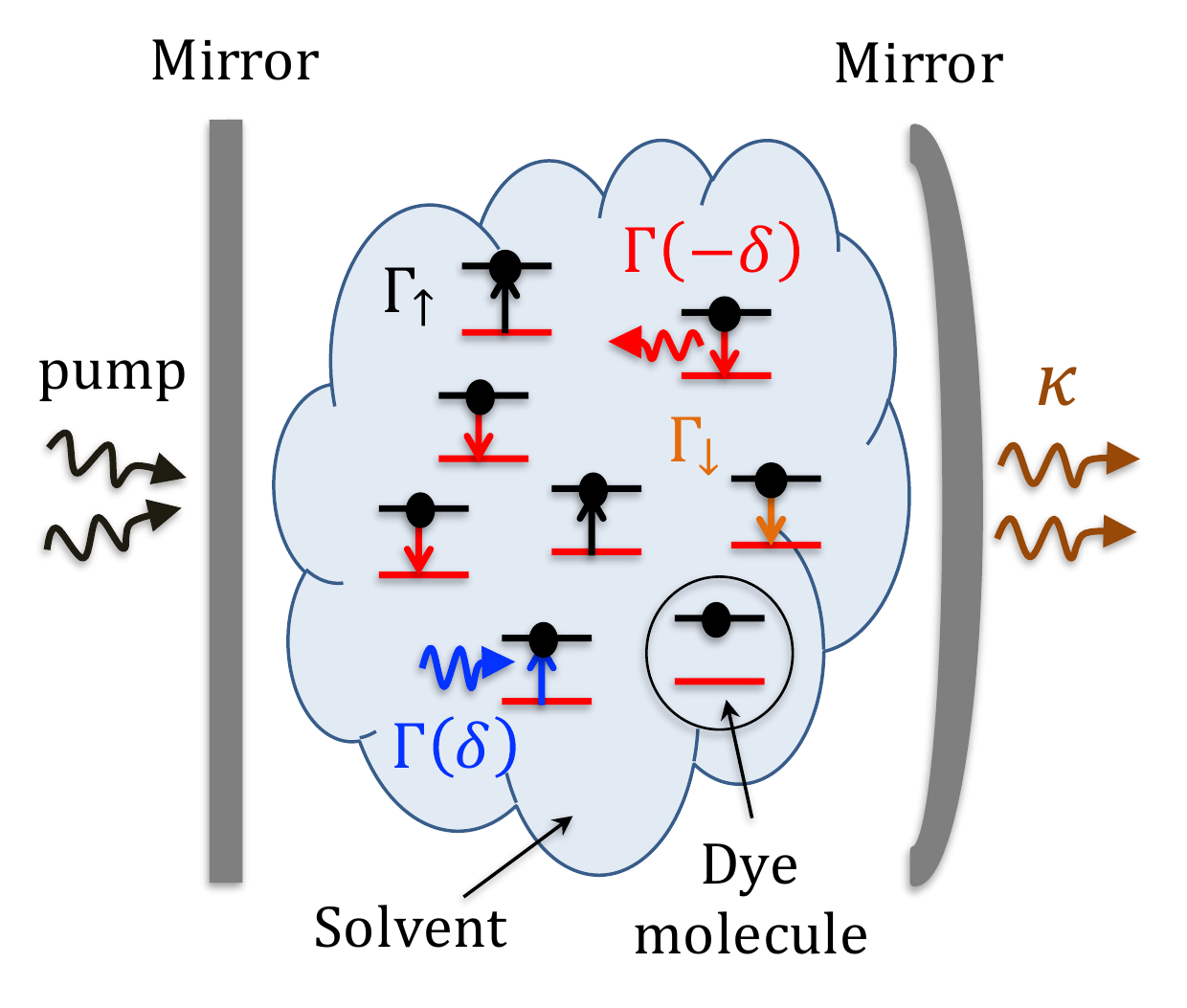}\hspace{15mm}
		\includegraphics[width=55mm]{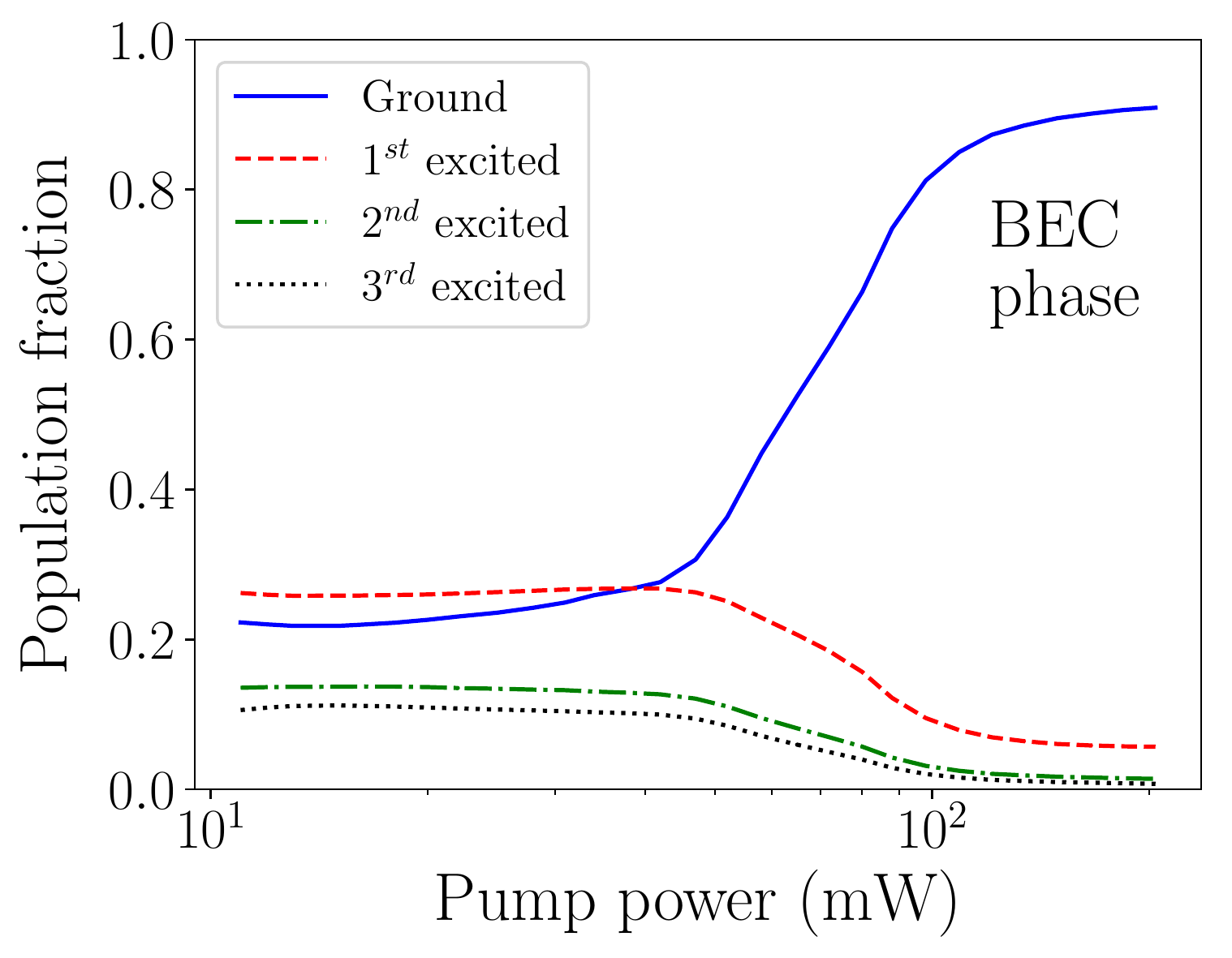}
	\vspace{-2ex}
		\caption{\label{fig:bec data}Formation of BEC inside a dye-filled open microcavity. (Left) A schematic of the system: the dye molecules in solution, are represented by two-level emitters and placed between two high-quality mirrors. Symbols are defined in the main text. (Right) Experimental data on the occupation of low-lying energy modes in a small photonic system (data re-purposed from  Ref.~\cite{Rodrigues20}). Beyond a critical pump power, a significant fraction of the total photons condense to the ground state. } 
\end{figure}	
\vspace{-2ex}
Several groups have achieved photon thermalisation and BEC using dye-filled open microcavities~\cite{Klaers10, Marelic15, Greveling18, Kassenberg20} as shown in \figref{fig:bec data}. An open microcavity consists of a pair of high-reflectivity, dielectric mirrors spaced a few half-wavelengths of light apart, at least one of which is curved. Light propagating has, in the paraxial approximation, a density of states exactly as described above, specifically equivalent to a two-dimensional harmonic oscillator (with DoS increasing linearly with energy above the ground state). The cavity is filled with a fluorescent dye such as Rhodamine~6G which is optically pumped. With a ground state energy of 2.1--2.2 electron-Volts (equivalent to 560-590 nm) the absorption by the dye is faster than cavity loss, so thermalisation occurs. For sufficiently strong pumping, BEC of photons is observed.


More recently, some other physical implementations of photon thermalisation and condensation have been developed. Plasmonic lattices can be used to enhance light-matter couplings with band-structures and materials suitable for BEC, as demonstrated by Hakala \textit{et al}~\cite{Hakala18}. Dopings in fibre lasers also have the appropriate relation between the absorption and emission spectra, and absorption is strong, so BEC is possible with appropriate intra-cavity filtering~\cite{Weill19}. Semiconductors built into microcavities are commonplace, and some evidence has come to light that pre-existing planar microcavity devices may show thermalisation and BEC~\cite{Barland19}.

As this article is intended as a conference proceedings, rather than reviewing the complete literature, we will now specialise to our own recent results, explaining the meaning rather than the details.

\subsection{Phase transitions in steady state}

Bose-Einstein condensation is an example of a phase transition. In the thermodynamic limit of a great many particles at thermal equilibrium, there are discontinuities in several properties as a function of the thermodynamic parameters (temperature and chemical potential). Photons thermalising in dye-filled microcavities constitute a finite-sized, driven-dissipative system which can, at best, only approximate thermal equilibrium. Therefore, not only are the phase transitions modified from the simple BEC case, but perhaps even the concept of phase transition should be re-evaluated.

\subsubsection{Breakdown of thermal equilibrium, including decondensation}

The thermalisation rate is effectively the rate at which photons are absorbed by the dye. This rate can be compared to the cavity photon loss rate to judge how nearly the system reaches thermal equilibrium. We made a numerical and theoretical study of the non-equilibrium phase diagram of a pumped dye-filled microcavity~\cite{Hesten18}. Steady-state photon populations were categorised as functions of pump and absorption rates relative to cavity loss rate. For the simulation parameters used, the population of a mode was found to jump by at least a factor 100 in a range of about 10\% variation in pump rate, which provided a working definition of a phase transition.

When absorption was at least 10 times faster than loss, the only phase transition detected was BEC. The opposite case (extremely weak absorption) lead to a series of phase transitions corresponding to single- or multi-mode lasing. Which modes transitioned depended sensitively on control conditions (pump spot size and position, thermalisation rate, pump rate). 

The intermediate, partially-thermalised case (with absorption and cavity loss rates being roughly equal) is perhaps the most interesting. As pump rate is increased, the first mode to condense (jump up in population) is robustly the ground state, but then further modes may condense. For very large pump rates, competition between modes for the gain given by the molecular excited state population leads to an effect we called \textit{decondensation}. In that case, the population of a mode decreased sharply with increasing pump rate.

\subsubsection{Fuzzy phase diagrams for small systems}

\begin{figure}[h]
\centering
\vspace{-2ex}
\includegraphics[width=0.65\textwidth]{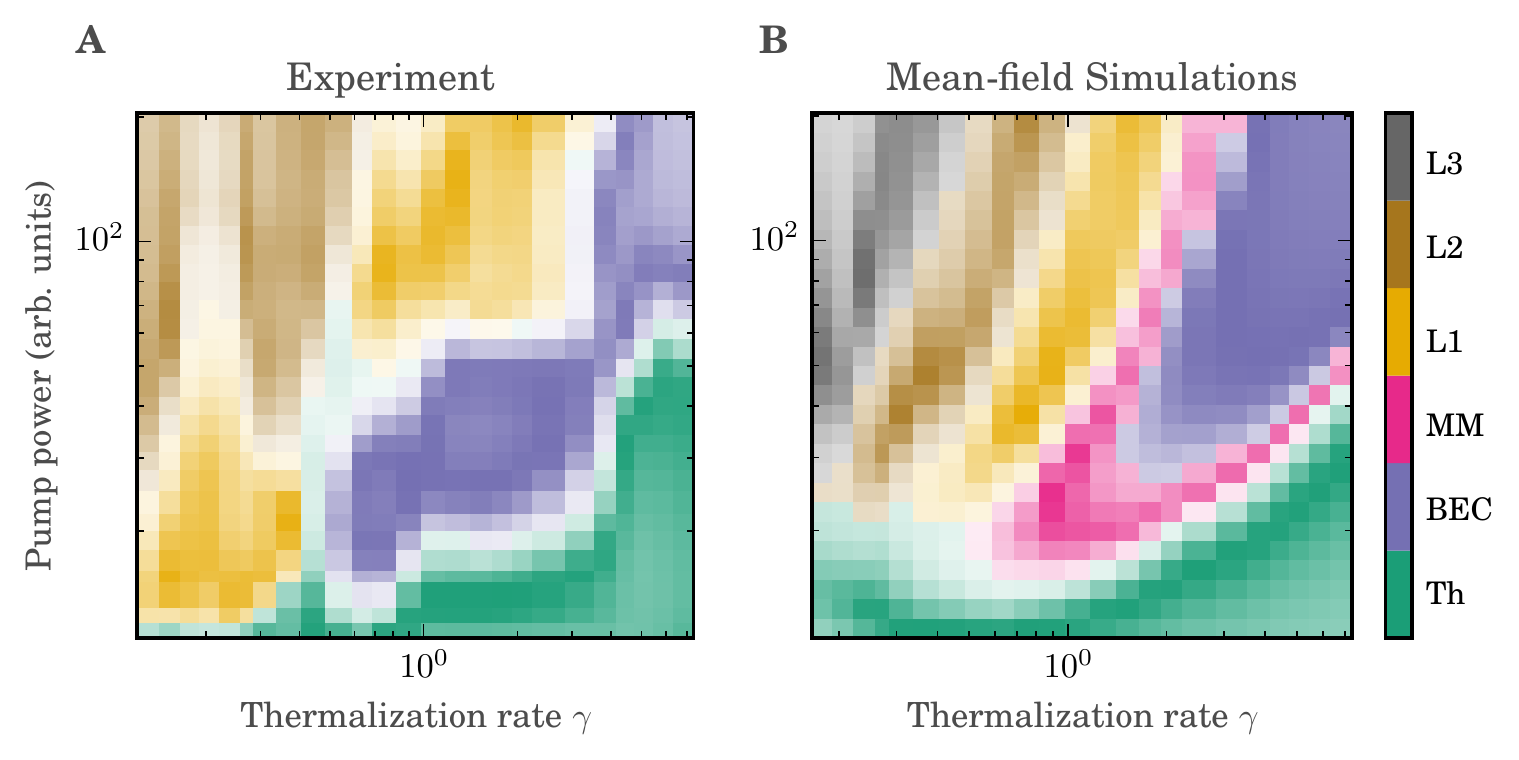}
\vspace{-2ex}
\caption{\label{fig:fuzzy phases}Fuzzy phases of photonic condensates. A: Experimental fuzzy phase diagram. By varying both the thermalization coefficient and the pump rate, our machine learning algorithm is able to identify the presence of several phases: thermal phase (Th); Bose-Einstein condensate (BEC); and two laser phases, characterized by the condensation of the first (L1) and the second (L2) excited states. B: The same, using mean-field, steady-state simulations, solved using a model implemented in the \pypbec{} package. The simulations help us interpret the experimental results. }
\end{figure}

We have subsequently explored the same phase diagram experimentally~\cite{Rodrigues20}: see \figref{fig:fuzzy phases}. Our apparatus allows us to measure the populations of individual energy levels, and we use those populations to assign phases to regions of experimental-parameter space (pump laser power and cavity length).

Unlike the simulations, the populations are small even at the threshold for condensation or lasing, with as few as 7 photons required for BEC~\cite{Walker18}. The phase transition is no longer a clear discontinuity in an order parameter, but rather is evidenced by slightly super-linear variation in population with respect to pump power. The phase transition cannot be assigned using heuristics (informed guess-work) so we turn to machine learning.

The populations over the parameter space show clustering, and we apply a fuzzy $c$-means algorithm to identify those clusters (the colours in \figref{fig:fuzzy phases}). Furthermore, we use a measure we call ``membership entropy'' to quantify the degree of affinity between population configurations and phases, i.e. how strongly we can say that a configuration sits within a phase. This measure is shown by the level of whiteness in the figure and represents the ambiguity in assigning configurations to phases. In this way, we construct a \textit{fuzzy phase diagram}. 

\vspace{-2ex}

\subsection{The dynamics of phase transitions}

We now turn to dynamical results (away from steady state), after quenches (theory work) and pulses (experiments). Close to phase transitions, the time taken to reach the steady-state behaviour increases dramatically, a phenomenon known as ``critical slowing down''. In simulations we observe exactly that behaviour~\cite{Walker19} for most of the BEC and multi-mode condensation transitions, except the decondensation. While the decondensation is associated with critical slowing down close to threshold, far above the threshold pump rate, there persists a slowing down by at least an order of magnitude with respect to the expected timescales of the system. This non-critical slowing is related to the strong gain competition which causes the decondensation.

\begin{figure}[h]
\centering
\vspace{-2ex}
\includegraphics[width=0.8\textwidth]{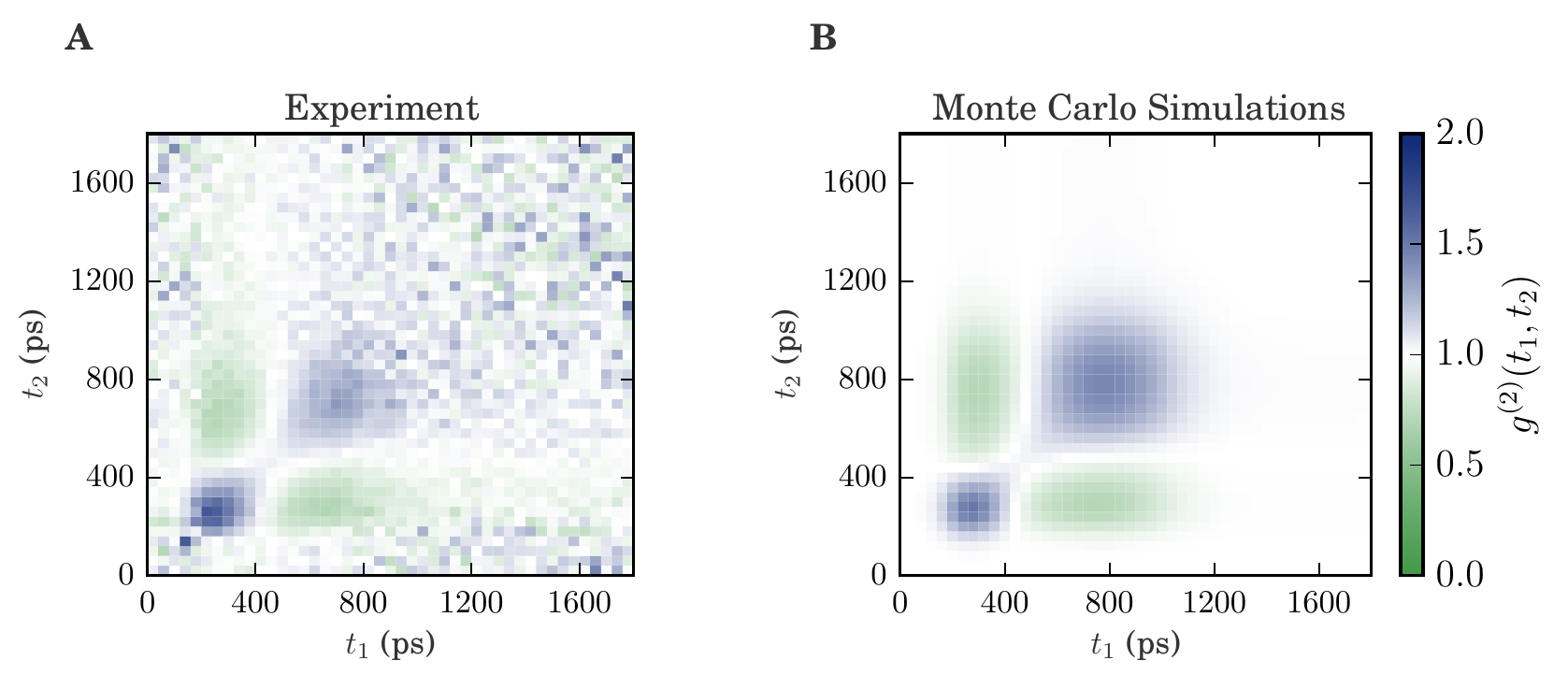}
\vspace{-2ex}
\caption{\label{fig: jitter}A: Experimental non-stationary second-order correlation function $g^{(2)}(t_1,t_2)$, clearly showing the presence of anti-correlation lobes, where $g^{(2)}<1$. These are a direct evidence of the presence of condensation jitter, and should not be mistaken with sub-Poissonian statistics. B: Results from a Monte Carlo simulation of the cavity dynamics, which can be  obtained from the \pypbec{} package, showing excellent agreement with the experiment.}
\vspace{-2ex}
\end{figure}

The dynamics in photon condensation normally happens in 5--500 ps. It is not feasible to quench a pump power between two values that fast, so we use 40-ps pulses to drive dynamics~\cite{Rodrigues20}. We then observe the dynamics using single-photon detectors (although operating in a regime of many photons)  and time-to-digital conversion electronics which have timing resolution as short as 40 ps. First, we observe the pulse-averaged behaviour to observe the total populations; we detect threshold behaviour as in the steady-state case. Then, we measure the average population as a function of time after the pulse (see \figref{fig: jitter}), and we detect the transient analogue of critical slowing down. Lastly, we use a non-stationary, two-time correlation measure $g^{(2)}(t_1, t_2)$. From the anti-correlations between early and late photon events, we note that the slowing down is related to strong jitter. The experiments match theory rather well. We conclude that the BEC phase transition is caused by (deterministic) stimulated scattering/emission but can only occur after (stochastic) spontaneous scattering of photons into the condensing mode.

\section{Generalised simulation of light in multi-mode resonators with fluorescent media}

Photon condensation is a rich phenomenon which is driven by both experiments and simulations. As such, it is important to have an accurate theory of the system based on a fundamentally sound physical model, with clear comparisons to experiments. We have such a model, which successfully predicts and explains many of our experimental observations~\cite{Rodrigues20, Walker18, Walker20, Marelic16}. We have shared our software for these simulations as an open-source project called \pypbec{} (\textit{Py}thon \textit{P}hoton \textit{B}ose-\textit{E}instein \textit{C}ondensation, available at \url{https://github.com/photonbec/PyPBEC}). Despite the name, it can be applied to a large range of physics involving multi-mode optical resonators interacting with fluorescent media, such as microlasers or plasmonic lattices.

\subsection{The model used and its limitations}

The model originates from work by Kirton and Keeling, where they start from a multi-mode, multi-molecule CQED framework, including the effects of mechanical relaxation (phonons) associated with the molecules~\cite{Kirton13, Kirton15}. They extended it to include effects of spatial inhomogeneities in the excitation of molecules~\cite{Keeling16}. The model is defined by the equations:
\begin{align}
\ddt{\hat{\rho}} =  -i[\hat{H}_0,\hat{\rho}] + \sum_{l,m} \left\{ \kappa \Lindblad[\hat{a}_m] + \Gup \Lindblad[\sigma^+_l] +  \Gdown \Lindblad[\sigma^-_l] + \Am \Lindblad[\hat{a}_m \sigma^+_l] + \Em \Lindblad[\hat{a}^\dagger_m \sigma^-_l]\right\} \hat{\rho},
\end{align}
where $\hat{\rho}$ is the density operator of the photon-molecule system. Here, $\hat{H}_0$ is the energy-conserving Hamiltonian and $\Lindblad[\hat{x}]\hat{\rho}$ = $\hat{x} \hat{\rho} \hat{x}^\dagger - \frac{1}{2} \left\{\hat{x}^\dagger \hat{x},\hat{\rho}\right\}$ is the standard Lindblad operator. Moreover, $\hat{a}_m$ and $\hat{a}^\dagger_m$ are the photon annihilation and creation operators for the cavity mode $m$, whereas $\sigma^\pm_l$ are the lowering (-) and raising (+) Pauli operators for the $l$th molecule. $A_m$ and $E_m$ are the absorption and emission rates for molecules into or from mode $m$, and \Gup{} is the rate of incoherent pumping of the molecules, $\kappa$ is the rate of photon loss and \Gdown{} is loss rate due to fluorescence in all non-cavity modes or other non-radiative processes.

In the form we use the equations most often (with a mean-field, semiclassical approximation assuming molecule and photon populations are uncorrelated), the above master equation can be used to derive a set of coupled equations of motion for the photon and molecular excitations:
\begin{eqnarray}
\ddt{n_m} &=&  - \kappa n_m + \sum_j \left\{E_m \,g_{m,j} M f_j(n_m+1) - A_m g_{m,j} M (1-f_j) n_m\right\}, \\
\ddt{f_j} &=& -\left\{\Gdown{} + \sum_m E_m g_{m,j}(n_m+1)\right\}f_j + \left\{\Gup{} + \sum_m A_m g_{m,j} n_m\right\}(1-f_j).
\end{eqnarray}
The molecules in the cavity have been divided into spatial bins $j$, with $f_j$ being the fraction of molecular excitations at $j$. The photon population in mode $m$ is given by $n_m$. Furthermore, $g_{m,j}$ is the coupling between the mode $m$ with molecules in the spatial bin $j$. We note that in the above equations we assume that the mode-mode coherence is weak and can be ignored if one is only interested in studying the photon and molecular excitations.

\subsection{The open-source code and its structure}

The code for simulating equations of motion is structured in three components:
\begin{itemize}
\item The photonic modes: the \texttt{Cavity} superclass defines all the energies and spatial representations of the resonator modes. It also defines the representation to be used if a spatially-resolved simulation is being used.
\item The fluorescent medium: the \texttt{OpticalMedium} class facilitates calculation of values for absorption and emission spectra at requested energies. A subclass of \texttt{OpticalMedium} called \texttt{Rhodamine6G} for obvious reasons is implemented, based on open-source data at \cite{RhodamineZenodo}. Emission and absorption rates can alternatively be input directly.
\item The solver (which defines and solves the equations of motion): the \texttt{Solver} superclass defines a format for recording the equations of motion, initial conditions, and the computational methods for solving the equations. Initially, three subclasses are defined, \texttt{SteadyState} for solving the equations above for the steady-state photon and molecule populations and \texttt{ODE} for the dynamics, as well as a \texttt{MonteCarlo} class which solves stochastically for the populations of the master equation without making the mean-field approximation.
\end{itemize}

The packages are written in \python {3}. A typical program starts by importing the three components. The \texttt{OpticalMedium} is either selected from the available list, custom defined, or emission and absorption rates at selected wavelengths may be input directly to the \texttt{Cavity} object. Once the \texttt{Cavity} is defined, the cavity parameters such as $\kappa$ and $g_{l,m}$ are set, including the rates of pumping $\Gup$. The \texttt{Solver} object is instantiated, and assigned a \texttt{Cavity} object and initial conditions. Finally the \texttt{Solver.solve} method is invoked, and results can be analysed and visualized.

\subsubsection{Notes on extensibility}
The code is structured so that the medium, the photonic modes, the coupling between medium and photonic modes, the equations of motion and the solution method can be easily customised. As such, it is extremely general. The specialisation to micro-cavity problems and photon BEC in particular comes about through the examples given, and the solvers, cavities and media which have been implemented.

At present, we have only implemented Rhodamine~6G as a medium, but other fluorescent dyes media based on many two-level emitters which undergo rapid dephasing can easily be added by including experimental data or simple theoretical models. The cavities can be any optical structure which exhibits discrete modes.

The choice of solver will of course play a significant role in both the results which can be obtained and the computational cost. The \texttt{MonteCarlo} solver runs much slower than the \texttt{SteadyState} solver, but gives beyond-mean-field results which match well to data, see Ref.~\cite{Walker20}. 

We plan to implement further solvers which take account of inter-mode phase coherences or use quantum-regression methods to efficiently calculate beyond-mean-field effects. Likewise, further materials and cavity types are also planned. 

\subsection{Example simulation: Stochastic dynamics of a two-mode laser}

\begin{figure}[h]
\centering
	\vspace{-2ex}
	\includegraphics[width=5in]{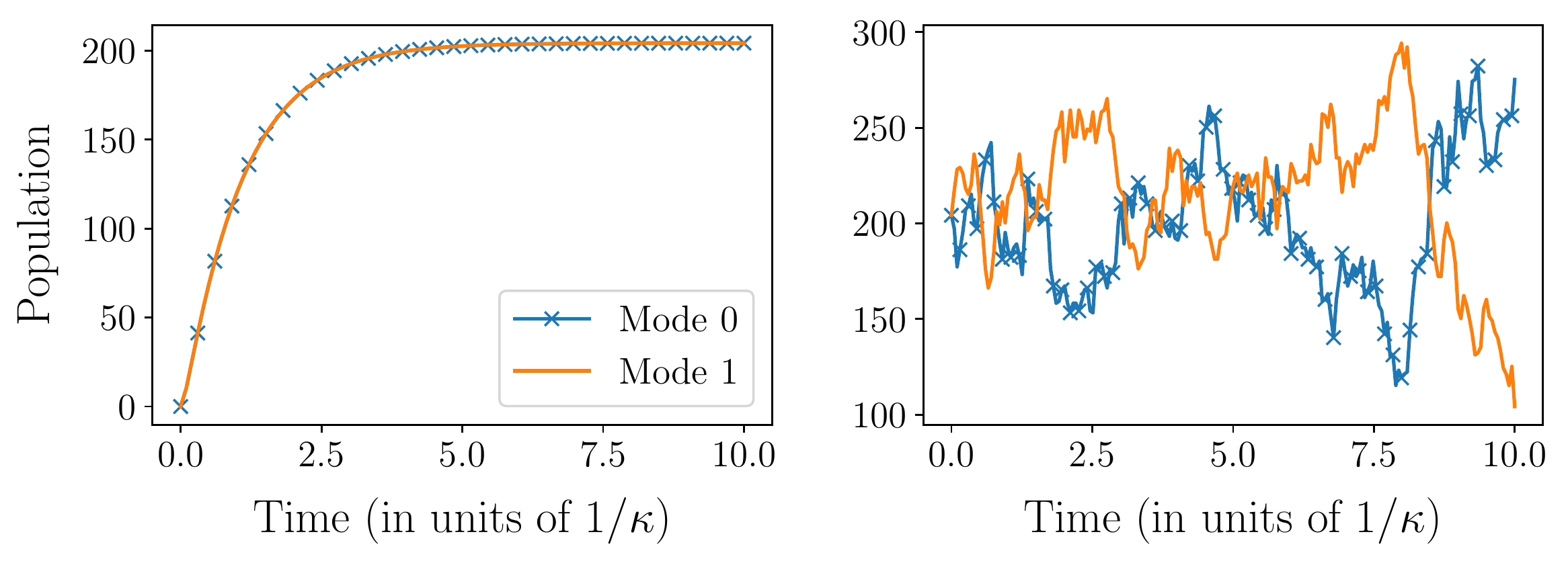}
	\vspace{-2ex}
	\caption{\label{fig: example simulation results}
	{The stochastic dynamics of the a pair of lasing modes. (Left) Mean-field evolution of the photon populations for modes 0 and 1, which are identical for the two degenerate modes. (Right) An example of stochastic realisation of the two populations starting from the steady state.}}
\end{figure}

{In this section, we give an example of how \pypbec{} can simulate the stochastic dynamics of a pair of degenerate lasing modes inside a dye-filled resonator. 
We use the cavity lifetime $1/\kappa$ as the unit of time, and  consider degenerate absorption and emission rates, $E_m$ = 50$\kappa$ and $A_m$ = 10$\kappa$ for $m=0,1$, with molecular pump and decay rates $\Gamma_{\uparrow},\Gamma_\downarrow$ = $\{100\kappa,\kappa\}$. The mean-field evolution of the two degenerate modes is obtained using the \texttt{ODE} solver to obtain the steady state of the system (\figref{fig: example simulation results}, left panel). The stochastic dynamics starting from the steady state are obtained with the \texttt{MonteCarlo} solver. \figref{fig: example simulation results}, right panel, shows how the photonic excitations switch between the two degenerate modes during one such stochastic evolution; populations are anticorrelated.}
 
{The code to generate the steady-state and stochastic dynamics shown in \figref{fig: example simulation results}, 
is given in the example folder of the \pypbec\ package. The high-level code is shown here. First, the \texttt{Cavity} class object is defined ($M$ and $J$ are the numbers of cavity modes and spatial bins respectively). }
\begin{small}
\begin{verbatim}
import numpy as np
from PyPBEC.Cavity import Cavity
cavity = Cavity(M=2, J=1, cavity_loss_rates=[1,1], cavity_emission_rates=[50,50], \
    cavity_absorption_rates=[1, 1], reservoir_decay_rates=[1], \
        reservoir_population=[1e9], coupling_terms=[[1],[1]])
cavity.set_reservoir_pump_rates(np.array([100]))
\end{verbatim}
\end{small}
An \texttt{OpticalMedium} object is not needed as we have explicitly defined absorption and emission rates. Then we solve the ordinary differential equation (ODE) to find mean-field evolution towards the steady state:
\begin{small}
\begin{verbatim}
from PyPBEC.Solver import ODE, MonteCarlo
solver_ode = ODE(cavity_obj=cavity, T=10, n_points=100)
solver_ode.set_initial_photons(np.array([0,0]))
solver_ode.set_initial_excited_molecules(np.array([0]))
solved_cavity_ode = solver_ode.solve()
\end{verbatim}
\end{small}
Lastly we solve the stochastic dynamics about the steady state
\begin{small}
\begin{verbatim}
solver_mc = MonteCarlo(cavity_obj=cavity, T=10, n_points=100)
solver_mc.set_initial_photons(solved_cavity_ode.photons[-1])
solver_mc.set_initial_excited_molecules(solved_cavity_ode.emols[-1])
solved_cavity_mc = solver_mc.solve()
\end{verbatim}
\end{small}
It is then a simple matter to plot the results, using the \texttt{.t} and \texttt{.photons} attributes of the resulting \texttt{Solver} objects.

\section{Discussion: how the simulations might be used}

We anticipate the package will not only be used in scientific research projects but also in teaching projects. For teaching reasons, clear documentation and a gallery of example programs are helpful. The repository comes with several examples in plain \python\ and in fully-annotated Jupyter notebooks. We will welcome community input in implementing features, and in supplying examples of simulations.

Photon BEC using dye-filled microcavities is not the only problem that can be solved using this package. Plasmonic lattices can readily be represented, as could a variety of resonators and metamaterials. Likewise, it is possible that modifications of the Kirton-Keeling model could be made to represent doped-glasses used for lasers, semiconductors, materials with optical-nonlinearities, or even gaseous media. A variety of non-equilibrium phenomena can be studied including fluctuations near criticality~\cite{Walker20}, transport~\cite{Dhar20} or emission of quantum-correlated light.

We thank Rupert Oulton, Peter Kirton and Tim Lappe for their contributions to this project, and acknowledge funding from EPSRC (UK) grant EP/S000755/1, and the European Commission via the PhoQuS project (H2020-FETFLAG-2018-03) number 820392.

\section*{References}

\bibliographystyle{iopart-num_NO_URL}
\bibliography{references}

\providecommand{\newblock}{}
\begin{thebibliography}{10}
\expandafter\ifx\csname url\endcsname\relax
  \def\url#1{{\tt #1}}\fi
\expandafter\ifx\csname urlprefix\endcsname\relax\def\urlprefix{URL }\fi
\providecommand{\eprint}[2][]{\url{#2}}

\bibitem{Christopoulos07}
Christopoulos S, von H\"ogersthal G~B~H, Grundy A~J~D, Lagoudakis P~G, Kavokin
  A~V, Baumberg J~J, Christmann G, Butt\'e R, Feltin E, Carlin J~F and
  Grandjean N 2007 {\em Phys. Rev. Lett.\/} {\bf 98}(12) 126405

\bibitem{KenaCohen10}
K{\'e}na-Cohen S and Forrest S 2010 {\em Nature Photonics\/} {\bf 4} 371

\bibitem{Bajoni08}
Bajoni D, Senellart P, Wertz E, Sagnes I, Miard A, Lema\^{\i}tre A and Bloch J
  2008 {\em Phys. Rev. Lett.\/} {\bf 100}(4) 047401

\bibitem{Kasprzak06}
Kasprzak J, Richard M, Kundermann S, Baas A, Jeambrun P, Keeling J, Marchetti
  F, Szyma{\'n}ska M, Andr{\'e} R, Staehli J {\em et~al.\/} 2006 {\em Nature\/}
  {\bf 443} 409--414

\bibitem{Plumhof14}
Plumhof J~D, St{\"o}ferle T, Mai L, Scherf U and Mahrt R~F 2014 {\em Nature
  materials\/} {\bf 13} 247--252

\bibitem{Daskalakis14}
Daskalakis K, Maier S, Murray R and K{\'e}na-Cohen S 2014 {\em Nature
  materials\/} {\bf 13} 271--278

\bibitem{Ramezani17}
Ramezani M, Halpin A, Fern{\'a}ndez-Dom{\'\i}nguez A~I, Feist J, Rodriguez
  S~R~K, Garcia-Vidal F~J and Rivas J~G 2017 {\em Optica\/} {\bf 4} 31--37

\bibitem{Klaers10}
Klaers J, Schmitt J, Vewinger F and Weitz M 2010 {\em Nature\/} {\bf 468}
  545--548 ISSN 1476-4687

\bibitem{Rodrigues20}
Rodrigues J~D, Dhar H~S, Walker B~T, Smith J~M, Oulton R~F, Mintert F and Nyman
  R~A 2020 {\em arXiv preprint arXiv:2006.12298\/}

\bibitem{Marelic15}
Marelic J and Nyman R~A 2015 {\em Phys. Rev. A\/} {\bf 91}(3) 033813

\bibitem{Greveling18}
Greveling S, Perrier K~L and van Oosten D 2018 {\em Phys. Rev. A\/} {\bf 98}(1)
  013810

\bibitem{Kassenberg20}
Kassenberg B, Vretenar M, Bissesar S and Klaers J 2020 {\em arXiv preprint
  arXiv:2001.09828\/}

\bibitem{Hakala18}
Hakala T~K, Moilanen A~J, V{\"a}kev{\"a}inen A~I, Guo R, Martikainen J~P,
  Daskalakis K~S, Rekola H~T, Julku A and T{\"o}rm{\"a} P 2018 {\em Nature
  Physics\/} {\bf 14} 739--744

\bibitem{Weill19}
Weill R, Bekker A, Levit B and Fischer B 2019 {\em Nature communications\/}
  {\bf 10} 1--6

\bibitem{Barland19}
Barland S, Azam P, Lippi G, Nyman R and Kaiser R 2019 {\em arXiv preprint
  arXiv:1912.11358\/}

\bibitem{Hesten18}
Hesten H~J, Nyman R~A and Mintert F 2018 {\em Phys. Rev. Lett.\/} {\bf 120}(4)
  040601

\bibitem{Walker18}
Walker B~T, Flatten L~C, Hesten H~J, Mintert F, Hunger D, Trichet A~A, Smith
  J~M and Nyman R~A 2018 {\em Nature Physics\/} {\bf 14} 1173--1177

\bibitem{Walker19}
Walker B~T, Hesten H~J, Dhar H~S, Nyman R~A and Mintert F 2019 {\em Phys. Rev.
  Lett.\/} {\bf 123}(20) 203602

\bibitem{Walker20}
Walker B~T, Rodrigues J~D, Dhar H~S, Oulton R~F, Mintert F and Nyman R~A 2020
  {\em Nature communications\/} {\bf 11} 1--11

\bibitem{Marelic16}
Marelic J, Zajiczek L~F, Hesten H~J, Leung K~H, Ong E~Y~X, Mintert F and Nyman
  R~A 2016 {\em New Journal of Physics\/} {\bf 18} 103012

\bibitem{Kirton13}
Kirton P and Keeling J 2013 {\em Phys. Rev. Lett.\/} {\bf 111}(10) 100404

\bibitem{Kirton15}
Kirton P and Keeling J 2015 {\em Phys. Rev. A\/} {\bf 91}(3) 033826

\bibitem{Keeling16}
Keeling J and Kirton P 2016 {\em Phys. Rev. A\/} {\bf 93}(1) 013829

\bibitem{RhodamineZenodo}
Nyman R~A 2017 {Absorption and Fluorescence spectra of Rhodamine 6G}

\bibitem{Dhar20}
Dhar H~S, Rodrigues J~D, Walker B~T, Oulton R~F, Nyman R~A and Mintert F 2020
  {\em arXiv preprint arXiv:2009.00094\/}

\end{thebibliography}

\end{document}